\documentclass[12pt]{article}
\usepackage{setspace}
\pdfoutput = 1
\usepackage[utf8]{inputenc}
\usepackage{dsfont}
\usepackage{color,graphicx}
\usepackage{epsfig}
\usepackage{amsmath}
\usepackage{verbatim}
\usepackage{amssymb}
\usepackage{mathabx}
\usepackage[mathcal]{eucal}
\usepackage{stmaryrd}
\usepackage{empheq}
\usepackage{esint}
\usepackage{physics}
\usepackage{amsfonts}
\usepackage{array}
\usepackage{setspace}
\usepackage{braket}
\usepackage{float}
\usepackage{url}
\usepackage{mathtools}
\usepackage{tensor}
\usepackage{bbold}
\usepackage{slashed}
\usepackage{tikz}
\usepackage{graphicx}
\usepackage{epstopdf}
\usepackage{subcaption}
\usepackage[labelfont=bf,font={sf}]{caption}
\usepackage{pgfplots}
\usepackage{mathrsfs}
\usepackage{fancybox}
\usepackage{eurosym}
\usepackage{tcolorbox}
\usepackage{tensor}
\usepackage[normalem]{ulem}
\allowdisplaybreaks
\usepackage{changepage}

\usepackage[margin = 1in]{geometry}
\usepackage[ragged]{footmisc}
\usepackage[bookmarks=true,bookmarksnumbered=false,
hyperindex=true,bookmarksopen=true,hyperfigures=true,
colorlinks=true,linkcolor=webblue,citecolor=webgreen,urlcolor=webblue,breaklinks]{hyperref}
\definecolor{webgreen}{rgb}{0, 0.5, 0}
\definecolor{webblue}{rgb}{0, 0, 0.5}
\definecolor{webred}{rgb}{0.5, 0, 0}
\definecolor{darkgreen}{rgb}{0,0.5,0}

\def\ben{\begin{equation}}
\def\een{\end{equation}}

     \let\r=v

\def\be{\begin{equation}}
\def\ee{\end{equation}}
\def\ba{\begin{array}}
\def\ea{\end{array}}

\def\dalemb#1#2{{\vbox{\hrule height .#2pt
 \hbox{\vrule width.#2pt height#1pt \kern#1pt
 \vrule width.#2pt}
 \hrule height.#2pt}}}

\newcommand{\bea}{\begin{eqnarray}}
\newcommand{\eea}{\end{eqnarray}}

\begin{document}

\thispagestyle{empty}
 ~\vspace{5mm}
\begin{adjustwidth}{-1cm}{-1cm}
\begin{center}
 {\Large \bf 
It costs nothing to teleport information into a black hole}
 \vspace{0.4in}

 {\bf Jonah Kudler-Flam${}^{1,2}$ and Geoff Penington${}^3$}
 \end{center}
 \end{adjustwidth}

\begin{center}
{March 31, 2025}

 \vspace{0.4in}
 {${}^1$School of Natural Sciences, Institute for Advanced Study, Princeton, NJ 08540, USA\\ ${}^2$Princeton Center for Theoretical Science, Princeton University, Princeton, NJ 08544, USA \\ ${}^3$Center for Theoretical Physics and Department of Physics,
University of California, Berkeley, California 94720, USA.}
 \vspace{0.1in}
 
 {\tt jkudlerflam@ias.edu, geoffp@berkeley.edu}
\end{center}

% \vspace{0.2in}

\begin{abstract}
\noindent
It is often claimed that adding a qubit to a black hole requires energy $\Delta E \geq T_H \log 2$ so that the extra Bekenstein-Hawking entropy can accommodate the qubit. In this essay, we explain how the recently discovered phenomenon of black hole decoherence allows quantum information to be teleported into a black hole, with arbitrarily small energy cost. The generalized second law is not violated and there is no conflict with unitarity because the teleportation creates new entanglement, analogous to Hawking radiation, between the black hole interior and exterior. In accordance with Landauer's principle, a nonzero minimum energy cost only appears when there is a net erasure of information and noise from the exterior or, equivalently, when ``zerobits'' are sent into the black hole.
\end{abstract}
\vspace{2.5 in}
{\small Essay written for the Gravity Research Foundation 2025 Awards for Essays on Gravitation}

%\tableofcontents
\pagebreak
\setcounter{page}{1}
\setcounter{tocdepth}{2}

%%%%%%%%%%%
% SECTION %
%%%%%%%%%%%
\section*{Introduction}

Common folklore says that we cannot add new quantum information to a black hole without also adding sufficient energy to account for the resulting increase in Bekenstein-Hawking entropy. In the simplest case, where a single qubit is added to the black hole, the minimum energy increase would be $T_{H} \log 2$ (with $T_H$ the Hawking temperature). 

A simple, if somewhat loose, argument for this is the following: assume the black hole is initially maximally entangled with a reference system $R$ and the qubit being sent in is maximally entangled with a second reference system $X$. If the process of sending information into the black hole is unitary, then it should preserve the entanglement entropy
\begin{align*}
 S(R X) = \frac{A_{\rm hor}}{4G} + \log 2.
\end{align*}
However, if this entropy is to be purified by the black hole $H$, we must have $S(H) = S(R X)$. This is only possible if the black hole has added enough energy to increase its thermodynamic entropy by at least one qubit. 

Similarly suggestive, but not quite airtight, arguments follow from the generalized second law (GSL), which says that any decrease in entropy of the exterior of a black hole must be compensated for by an increase in its horizon area, and the closely related Bekenstein bound, which says that the entropy of any quantum system is upper bounded by $2\pi$ times the product of its energy and radial size. Intuitively, sending a qubit into a black hole can remove one bit of entropy from the exterior, violating the GSL unless the horizon area increases. Similarly, for a quantum system to fall into the black hole its radial size must be small enough to fit into the black hole horizon; a small amount of work then shows that the Bekenstein bound requires the black hole mass to increase by $\delta M \geq T_H \log 2$.

The purpose of the present essay is to explain that, while the generalized second law and Bekenstein bound are certainly true, the folklore that adding information to a black hole necessarily carries an energy cost is not. Our counterexample to this claim will be built on a recent gedankenexperiment by Danielson, Satishchandran, and Wald (DSW) \cite{2022IJMPD..3141003D} where an experimentalist, Alice, sends a charged particle, initially with spin aligned in the $x$-direction, through a Stern-Gerlach apparatus to spatially separate the branches of its wavefunction
\begin{align*}
 \ket{\Psi} = \frac{1}{\sqrt{2}}\left(\ket{\uparrow}\ket{\psi_{\uparrow}} + \ket{\downarrow}\ket{\psi_{\downarrow}}\right),
\end{align*}
where $\psi_{\uparrow,\downarrow}$ are coherent states of the electromagnetic field sourced by the spin following the respective path of its branch of the wavefunction.
After waiting for proper time $T$, she recombines the wavefunction, as shown in Fig. \ref{fig:DSW_tikz} (left). Her density matrix is then
\begin{align}\label{eq:densitydecohereintro}
 \rho= \frac{1}{2}\begin{pmatrix}
 1 & \bra{\psi_{\uparrow}}\psi_{\downarrow}\rangle \\
 \bra{\psi_{\downarrow}}\psi_{\uparrow}\rangle & 1
 \end{pmatrix}.
\end{align}
In a flat spacetime, the radiation from the particle can be made arbitrarily small by making the Stern-Gerlach apparatus sufficiently adiabatic. Assuming no observer attempts to measure the electric field, the quantum coherence will then be retained so that the off-diagonal terms in \eqref{eq:densitydecohereintro} will be close to unity. However, in the presence of a Killing horizon, such as a black hole, DSW showed that decoherence is inevitable, with the off-diagonals decaying exponentially in $T$ \cite{2023PhRvD.108b5007D}.\footnote{A similar effect occurs for neutral particles, but the decoherence is much slower because it is mediated by gravity rather than electromagnetism.} This is due to low-frequency ``soft'' radiation emitted from Alice's particle into the black hole. The black hole harvests ``which-path'' information about Alice's particle and this information cannot be coherently recovered by the particle during the recombination process (as it would be in flat space) because of the presence of the horizon. Crucially, because no ``hard'' radiation is produced, the energy added to the black hole during this process vanishes in the adiabatic limit. 

%In Section \ref{sec:teleport}, we explain how the DSW decoherence protocol can be used to teleport quantum information into a black hole with negligible energy cost. In Section \ref{sec:page}, we explain how this result is consistent with unitary evolution of black hole microstates. Finally, in Section \ref{sec:zerobits}, we leverage the language of quantum communication resource theories to the communication protocols that are possible without adding energy to the black hole. The resulting theory is a novel variant of the usual resource theory of quantum communication and can be thought of as the ``Stinespring complement'' of the resource theory of purity (or, more generally, of quantum thermodynamics). \JKF{citations?}

\section*{Teleportation by black hole decoherence}\label{sec:teleport}

\begin{figure}
 \centering
 \begin{tikzpicture}[scale = 1]
    \draw[thick] (2,2) -- (5,5);
    \draw[double]  (8.5,1.5) -- (5,5);
    \draw[thick] (5.5-.5,0) to[out=90, in = -90]  (5.5-.5,1.5)to[out=90, in = -90] (5-.5,2)to[out=90, in = -90] (5-.5,3) to[out=90, in = -90]  (5.5-.5,3.5)to[out=90, in = -90] (5.5-.5,5);
    \draw[thick,dashed] (5.5-.5,.5) to[out=90, in = -90] (5.5-.5,1.5)to[out=90, in = -90] (6-.5,2)to[out=90, in = -90] (6-.5,3) to[out=90, in = -90] (5.5-.5,3.5)to[out=90, in = -90] (5.5-.5,4.5);
    \draw[|-| ,thick] (6.5-.5,1.8) -- (6.5-.5,3.2);
    \node[] at (6.75-.5,2.5) {$T$};
    \node[red] at (2.75,4) {$B_1$};
    \node[] at (6-.75,4) {$A$};
      \draw[decorate, decoration={zigzag}] (2,5) -- (5,5);
      \draw[decorate, red, ->,decoration={snake,amplitude=1mm, segment length=12mm}] (4,2.5) -- (3,3.5);
      \draw[decorate, red, ->,decoration={snake,amplitude=1mm, segment length=12mm}] (4.25,2.75) -- (3.25,3.75);
  \filldraw[black] (5,5) circle (1.5pt);
  \node[] at (4.4,2.5) {$\uparrow$};
   \node[] at (5.4,2.5) {$\downarrow$};
   \node[] at (5,1) {$\rightarrow$};
\end{tikzpicture}
 ~~~~~~
 \begin{tikzpicture}[scale = 1]
    \draw[thick] (2,2) -- (5,5);
    \draw[double]  (8.5,1.5) -- (5,5);
    \draw[thick,dashed] (5,.5) to[out=90, in = -90] (5,1)to[out=90, in = -90] (5.25,1.25)to[out=90, in = -90] (5.25,1.75) to[out=90, in = -90] (5,2)to[out=90, in = -90] (5,2.5)to[out=90, in = -90] (5.25,2.75)to[out=90, in = -90] (5.25,3.25)to[out=90, in = -90] (5.,3.5)to[out=90, in = -90] (5,4.5);    
        \draw[thick] (5,0) to[out=90, in = -90] (5,1)to[out=90, in = -90] (4.75,1.25)to[out=90, in = -90] (4.75,1.75) to[out=90, in = -90] (5,2)to[out=90, in = -90] (5,2.5)to[out=90, in = -90] (4.75,2.75)to[out=90, in = -90] (4.75,3.25)to[out=90, in = -90] (5.,3.5)to[out=90, in = -90] (5,5);   
    \node[red] at (2.75,3.25) {$B_1$};
    \node[red] at (3.15,4.35) {$B_2$};
    \node[] at (6-.75,4) {$A$};
      \draw[decorate, decoration={zigzag}] (2,5) -- (5,5);
      \draw[decorate, red, ->,decoration={snake,amplitude=1mm, segment length=12mm}] (4.25,1.5) -- (3.25,2.5);
      \draw[decorate, red,->, decoration={snake,amplitude=1mm, segment length=12mm}] (4.5,1.75) -- (3.5,2.75);
            \draw[decorate, red, ->,decoration={snake,amplitude=1mm, segment length=12mm}] (4.25,3) -- (3.25,4);
      \draw[decorate, red,->, decoration={snake,amplitude=1mm, segment length=12mm}] (4.5,3.25) -- (3.5,4.25);
  \filldraw[black] (5,5) circle (1.5pt);
  \draw[draw=black, fill=white] (4.85,2.45) rectangle (5.15,2.05);
    \node at (5, 2.25) {\tiny \textbf{H}};
    \node[] at (5.35,0.25) {$\ket{\psi}$};
\end{tikzpicture}
 \caption{Left: The DSW protocol with Alice, $A$, decohering via emission of soft quanta, $B_1$, into the black hole. Right: Our teleportation protocol where Alice implements the DSW protocol two times, with the Hadamard, $H$, implemented between to rotate bases.}
 \label{fig:DSW_tikz}
\end{figure}
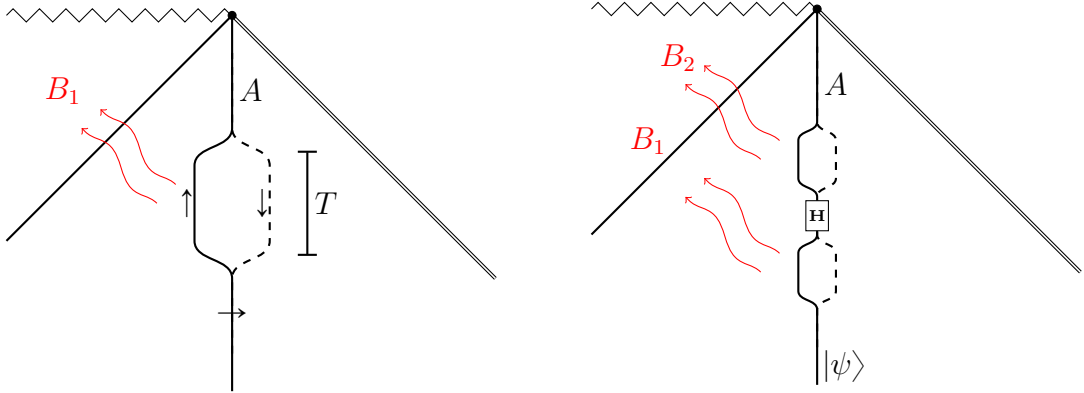

From the perspective of an exterior observer, the late-time limit of the DSW decoherence protocol is described by the dephasing channel
\begin{align*}
 \mathcal{N}\begin{pmatrix}
 \rho_{00} & \rho_{01} \\
 \rho_{10} & \rho_{11}
 \end{pmatrix} = \begin{pmatrix}
 \rho_{00} & 0 \\
 0 & \rho_{11}
 \end{pmatrix}
\end{align*}
that kills all off-diagonal elements of the density matrix in the $\ket{\uparrow} = \ket{0},\ket{\downarrow} = \ket{1}$ basis. Like all quantum channels, this can be extended to a Stinespring dilation
\begin{align}\label{eq:cobit}
 V \left(\alpha \ket{0}_A + \beta \ket{1}_A\right) = \alpha \ket{0}_A \ket{0}_{B_1} + \beta \ket{1}_A \ket{1}_{B_1},
\end{align}
where $B_1$ is a purifying reference system that is traced over to recover the original channel. In this case, $B_1$ describes soft radiation in the interior of the black hole, as shown in Figure \ref{fig:DSW_tikz} \cite{2025arXiv250104773D}. The exterior forgets the relative phase between $\ket{0}$ and $\ket{1}$ because the state is being measured by the interior in the $\ket{0},\ket{1}$ basis. 

A somewhat trivial, but important, observation is that we can use the DSW protocol to send classical information into the black hole without adding energy to it. Depending on the state of the classical bit that we want to send, we prepare the particle in either the state $\ket{\uparrow}$ or $\ket{\downarrow}$. By measuring the which-path information, an observer (or, if necessary, a collection of observers) in the interior can recover the classical bit. 

However, we can also do better than this. The isometry $V$ given in \eqref{eq:cobit} defines the communication of a coherent bit or ``cobit'' channel from Alice (in the exterior) to Bob (in the interior) \cite{2004PhRvL..92i7902H}. As we just saw, cobits can be used to communicate classical bits by encoding the classical bit into computational basis states. However, when the input state is a superposition of computational basis states (as in the original DSW protocol) they can also create entanglement between Alice and Bob. As a result, cobits are sufficient to teleport quantum information into the black hole.

% \begin{figure}
% \centering
% \input{teleport_tikz}
% \caption{Caption}
% \label{fig:enter-label}
% \end{figure}
Let us describe explicitly how such a teleportation works. Alice encodes the qubit
\begin{align*}
 \ket{\psi} = \alpha \ket{0} + \beta\ket{1}
\end{align*}
in the spin state of the charged particle. She then carries out the DSW decoherence protocol, producing the state \eqref{eq:cobit}. However, after recombining the wavefunction, Alice can apply a Hadamard gate $H_A$ by rotating the particle spin. This produces the state
\begin{align*}
 H_A V_1 \ket{\psi} = \alpha \ket{+}_A\ket{0}_{B_1} + \beta\ket{-}_A\ket{1}_{B_1}
\end{align*}
where $\ket{\pm} = 1/\sqrt{2} \left(\ket{0} \pm \ket{1}\right)$. Finally, she repeats the DSW decoherence protocol, as shown in Fig. \ref{fig:DSW_tikz} (right), producing the state
\begin{align*}
 V_2 H_A V_1 \ket{\psi} = \frac{1}{\sqrt{2}}\ket{0}_A \left(\alpha\ket{0}_{B_1}\ket{0}_{B_2}+\beta\ket{1}_{B_1}\ket{0}_{B_2}\right) + \frac{1}{\sqrt{2}}\ket{1}_A \left(\alpha\ket{0}_{B_1}\ket{1}_{B_2}-\beta\ket{1}_{B_1}\ket{1}_{B_2}\right).
\end{align*}
The system $B_2$ is again soft radiation inside the black hole. From Alice's point of view, the state has now completely decohered; her density matrix is maximally mixed. As a result, all information about the state $\ket{\psi}$ can be recovered from inside the black hole. In fact, if Bob, in the black hole interior, applies a Controlled $Z$ gate to the two qubits $B_1$ and $B_2$ encoded in the soft radiation, we obtain the state 
\begin{align}\label{eq:finalteleportation}
 CZ_{B_1 B_2} V_2 H_A V_1 \ket{\psi}=\frac{1}{\sqrt{2}}\left(\ket{0}_A \ket{0}_{B_2}+ \ket{1}_A\ket{ 1}_{B_2}\right) \ket{\psi}_{B_1}.
\end{align}
Alice's initial state has been teleported to Bob's first ``register'' $B_1$, while the second register $B_2$ has become maximally entangled with the spin of Alice's particle.

\section*{Unitarity and the information problem}\label{sec:page}

How was this protocol able to avoid our arguments in the introduction that sending information into a black hole should always add energy to it? The crucial ingredient is the Bell pair shared between $A$ and $B_2$ in \eqref{eq:finalteleportation}. This is new entanglement between the black hole interior and exterior that was created by the teleportation process. As a result, the generalized second law is not violated: the increase in exterior entropy from the new entanglement is always at least as large as the decrease from sending the qubit into the black hole. Thus, there is no need for the horizon area to increase. In essence, the Bell pair created during the teleportation acts as a sort of synthetic Hawking quantum that prevents a violation of the GSL, just like the entropy of actual Hawking radiation prevents a violation of the GSL when horizon area decreases during black hole evaporation. Indeed, it has recently been argued that, in the charged particle's frame of reference, DSW decoherence results directly from the interaction of the particle with ``soft Hawking radiation'' \cite{2024PhRvD.110d5002W, 2024arXiv240502227B,2025PhRvD.111b5014D}.

What about our argument based on unitarity of the microscopic black hole dynamics? Consider a black hole that has been evaporating for longer than the Page time so that it is maximally entangled with a radiation system $R$ in the state
\begin{align}\label{eq:psimaxentangled}
 \ket{\Phi}_{HR} = e^{-S_{BH}/2}\sum_i \ket{\phi_i}_{H}\ket{i}_R
\end{align}
where $\ket{\phi_i}_H$ labels an orthonormal basis of $e^{S_{BH}}$ black hole microstates and $\ket{i}_R$ are orthogonal states of the radiation Hilbert space, which has dimension much larger than $e^{S_{BH}}$ because we are past the Page time. 

Let us try to understand what happens when we implement our decoherence teleportation protocol. We can model the microscopic decoherence process by a unitary
\begin{align*}
U_{\rm decohere} = \ket{0}\!\bra{0}_A \otimes U_0 + \ket{1}\!\bra{1}_A \otimes U_1
\end{align*}
so that the states $\ket{0}, \ket{1}$ of the qubit $A$ are preserved, but affect the dynamics $U_0, U_1$ of the black hole. For example, if the black hole starts in some fixed microstate $\ket{\phi_i}_B$, we obtain
\begin{align*}
U_{\rm decohere} \ket{\psi}_A \ket{\phi_i}_H = \alpha \ket{0}_A U_0 \ket{\phi_i}_H + \beta \ket{1}_A U_1 \ket{\phi_i}_H.
\end{align*}
Generically, at late times, we should expect $U_0 \ket{\phi_i}_H$ and $U_1 \ket{\phi_i}_H$ to be approximately orthogonal and so the state of $A$ dephases, in agreement with the semiclassical picture. If we then applied a Hadamard gate and allow the qubit $A$ to decohere a second time (as in our qubit teleportation protocol), the state of $A$ would end up maximally mixed.

Now, suppose instead that the black hole starts in the entangled state $\ket{\Phi}_{HR}$ defined in \eqref{eq:psimaxentangled}. After implementing the teleportation protocol, we get
\begin{align}\label{eq:microdecohere}
\begin{aligned}
 U_{\rm decohere} H_A U_{\rm decohere} \ket{\psi}_A\ket{\Phi}_{HR}=& \ket{0}_A e^{-S_{BH}/2}\sum_i\left(\alpha U_0 U_0 + \beta U_0 U_1\right) \ket{\phi_i}_H\ket{i}_R \\&\qquad+ \ket{1}_A e^{-S_{BH}/2}\sum_i\left(\alpha U_1 U_0 - \beta U_1 U_1\right) \ket{\phi_i}_H\ket{i}_R. 
\end{aligned}
\end{align}
Generically, the states $U_aU_b \ket{\phi_i}_B$ and $U_{a'}U_{b'} \ket{\phi_j}_H$ will be approximately orthogonal whenever $a \neq a'$, $b \neq b'$, or $i \neq j$. Consequently, the original qubit $A$ will end up very close to maximally mixed, as expected from the semiclassical calculation. However, the state $\ket{\psi}$ has no longer been teleported into the microscopic black hole degrees of freedom $H$; indeed a simple calculation shows that $H$ is also maximally mixed. Instead, the three quantum subsystems $A$, $H$, and $R$ form a quantum erasure code; on their own, each subsystem carries no information about $\ket{\psi}$. As a result, all information about $\ket{\psi}$ can be recovered from \textit{any} two of the three subsystems. 

In particular, $\ket{\psi}$ can be recovered from a combination of the radiation $R$ and the decohered spin $A$. To see this, suppose, as in the introduction, that $A$ started in a maximally entangled state $\ket{\Psi}_{AX} = \tfrac{1}{\sqrt{2}}\left(\ket{00} + \ket{11}\right)$ purified by yet another auxiliary system $X$. By unitarity, the entropy $S(RX)$ must be conserved and equal to $\left(S_{BH} + \log 2\right)$. Despite what we naively suggested in the introduction, this is not paradoxical because $RX$ is purified not just by the black hole $H$ but also by $A$. However, it does have to be the case that $S(RAX) = S(H) \leq S_{BH}$. We have $S(RA) =S(HX) = S_{BH} + \log 2$, since we know from \eqref{eq:microdecohere} that $H$ is maximally mixed for any fixed state $\ket{\psi}$. It follows that $X$ is maximally entangled with $RA$ and, consequently, that the latter encodes all quantum information originally in $A$.

\begin{figure}
 \centering
 \begin{tikzpicture}[scale = 1]
    \draw[thick] (1.5,1.5) -- (5,5);
    \draw[double]  (8.5,1.5) -- (5,5);
    \filldraw[lightgray,opacity = .5] (1.5,5) -- (3.15,3.35) -- (1.5, 3.35-1.65);
    \draw[thick,dashed] (5,.5) to[out=90, in = -90] (5,1.25)to[out=90, in = -90] (5.25,1.5)to[out=90, in = -90] (5.25,1.875) to[out=90, in = -90] (5,2.125)to[out=90, in = -90] (5,2.625)to[out=90, in = -90] (5.25,2.875)to[out=90, in = -90] (5.25,3.25)to[out=90, in = -90] (5.,3.5)to[out=90, in = -90] (5,4.5);    
    \draw[thick] (5,.5) to[out=90, in = -90] (5,1.25)to[out=90, in = -90] (4.75,1.5)to[out=90, in = -90] (4.75,1.875) to[out=90, in = -90] (5,2.125)to[out=90, in = -90] (5,2.625)to[out=90, in = -90] (4.75,2.875)to[out=90, in = -90] (4.75,3.25)to[out=90, in = -90] (5.,3.5)to[out=90, in = -90] (5,5);   
    \node[darkgreen] at (2,3.5) {$B_i$};
    \node[red] at (3.25,4.25) {$B_o$};
    \node[] at (6-.75,4) {$A$};
      \draw[decorate, decoration={zigzag}] (1.5,5) -- (5,5);
      \draw[decorate, red,->, decoration={snake,amplitude=1mm, segment length=12mm}] (2.75+.5,1) -- (3.75+.5,2);
        \draw[decorate, darkgreen, ->,decoration={snake,amplitude=1mm, segment length=12mm}] (1.5,2.5) -- (2.5,3.5);
         \draw[decorate, darkgreen, ->,decoration={snake,amplitude=1mm, segment length=12mm}] (1.75,2.25) -- (2.75,3.25);
      \draw[decorate, red,->, decoration={snake,amplitude=1mm, segment length=12mm}] (2.5+.5,1.25) -- (3.5+.5,2.25);
    \draw[decorate, red, ->,decoration={snake,amplitude=1mm, segment length=12mm}] (4.25,2.675) -- (3.25,3.675);
      \draw[decorate, red,->, decoration={snake,amplitude=1mm, segment length=12mm}] (4.5,3-.125) -- (3.5,4-.125);
      \draw[decorate, blue,->, decoration={snake,amplitude=1mm, segment length=2mm}] (6.75,.75) -- (7.75,1.75);
       \draw[decorate, blue,->, decoration={snake,amplitude=1mm, segment length=2mm}] (7,0.5) -- (8,1.5);
       \node[blue] at (7,1.5) {$R$};
  \filldraw[black] (5,5) circle (1.5pt);
\end{tikzpicture}
 \caption{After the Page time, the Hawking radiation $R$ (blue) of an old black hole has an island (gray). Alice's particle interacts with soft Hawking quanta (red) which are themselves entangled with modes in the island (green).
 }
 \label{fig:enter-label}
\end{figure}
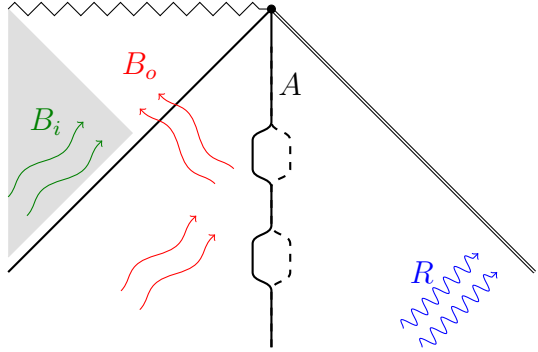

How is this microscopic story consistent with the semiclassical picture, where all information about the state $\ket{\psi}$ ends up inside the black hole? There is a close analogy between the recovery of $\ket{\psi}$ from $RA$ and the Hayden-Preskill process \cite{2007JHEP...09..120H}, where a small quantum system, or diary, thrown into a black hole after the Page time, can be recovered from the radiation by collecting a few extra Hawking quanta after waiting for a scrambling time $(2\pi T_H)^{-1}\log S_{BH}$. (These extra quanta play the role of the spin $A$.) In both cases, information that is semiclassically localised in the interior can somehow be recovered from spacelike-separated radiation in the exterior. The semiclassical and microscopic pictures of the Hayden-Preskill process have recently been reconciled by the observation that, after the Page time, the Hawking radiation has an ``entanglement island'' that includes much of the black hole interior \cite{2020JHEP...09..002P,2019JHEP...12..063A}. Nonperturbative quantum gravity effects mean that any information within this island can be secretly recovered by using exponentially complex operators acting on the radiation. (One way to see these effects is through a replica trick calculation of the recovery process, where they manifest as the inclusion of replica wormholes in the gravitational path integral \cite{2020JHEP...05..013A, 2019arXiv191111977P}.) The island of the early radiation does not include any part of the worldline of the diary, and so (as expected) the early radiation does not know about its state. However, the inclusion of the additional late radiation enlarges the associated island. Consistent with the Hayden-Preskill prediction, the island includes the diary, which can then be recovered, if and only if some of the radiation was emitted more than a scrambling time after the diary fell in.

The case of information teleported into a black hole via decoherence is similar but subtly different. Because there is no scrambling time delay, adding the additional decohered qubit $A$ does not significantly enlarge the island associated to $R$. So how can $\ket{\psi}$ be encoded in $AR$, but not in $R$ alone? The answer is that even in the semiclassical description the teleported information $\ket{\psi}$ is not localised on any particular worldline in spacetime. In particular, semiclassical causality ensures that all soft modes $B_o$ directly interacting with $A$ are outside the island (even though they eventually fall into the black hole). However those modes are highly entangled with other soft modes $B_i$ that are inside the island\footnote{The tensor product decomposition of the soft radiation $B$ as $B_i \otimes B_o$ should not be confused with our earlier decomposition $B_1 \otimes B_2$. The former separates degrees of freedom by their physical location within the interior, while the latter separated degrees of freedom by the decoherence process that entangled them with $A$.} (see Figure \ref{fig:enter-label}). As a result, the state $\ket{\psi}$ ends up encoded in a tripartite entangled state of $A$, $B_o$, and $B_i$. In fact, this tripartite semiclassical state forms exactly the same sort of quantum erasure code that we previously claimed should exist microscopically for $A$, $H$, and $R$. The state $\ket{\psi}$ ends up localised in the interior, in the sense that it can be recovered from a combination of $B_i$ and $B_o$. But it can also be recovered from either $A B_o$ or $AB_i$. Since the teleported information can be recovered semiclassically from a combination of the island and the decohered spin $A$ (but not from the island alone), it can also be recovered nonperturbatively from a combination of $R$ and $A$ (but not from $R$ alone), even though $A$ has negligible effect on the size of the island. 

\section*{Zerobits cost energy; ebits release energy}\label{sec:zerobits}

A powerful way to characterize quantum communication protocols is through the framework of resource theories. For example, the teleportation by decoherence protocol used two cobit channels to send a qubit into the black hole, while creating a Bell pair (or ``ebit'') of entanglement between interior and exterior. We can write this as the resource inequality
\begin{align*}
% \label{eq:cobitinequality}
 \text{2 cobits}~~~ \geq~~~ \text{1 qubit + 1 ebit}.
\end{align*}
In fact, this inequality also holds with the sign reversed, so that it is really a resource equality
\begin{align}\label{eq:cobitequality}
 \text{2 cobits}~~~ =~~~ \text{1 qubit + 1 ebit}.
\end{align}

In the conventional resource theory of quantum communication, the ebit on the right-hand side of \eqref{eq:cobitequality} is a positive, valuable resource. Bell pairs can be a useful asset in e.g. teleportation protocols using only classical communication. If the ebit is not needed, then it can always be thrown away.

When communicating from the exterior into the interior of a black hole, however, the opposite is true. Entanglement between the interior and exterior can be created for ``free'' (in fact at a negative energy cost) by simply allowing the black hole to radiate. Thanks to the generalized second law, however, this entanglement cannot be destroyed without paying a price in energy. Ebits are therefore a negative resource, or cost, that needs to be paid to send quantum information using cobit teleportation. This is why cobits can be sent into a black hole with negligible energy cost, while sending qubits in without creating entanglement always costs energy (because otherwise any such protocol could be used to violate the GSL). 

We have seen that a qubit (with no additional ebits) costs energy $T_H \log 2$, a cobit costs negligible energy, and an ebit releases energy $T_H \log 2$. Any method of sending a qubit can also be used to send a cobit by preparing the appropriate state in Alice's lab and then sending Bob's qubit to him, so
\begin{align*}
 \text{1 qubit} ~~~\geq ~~~ \text{1 cobit}.
\end{align*}
Somehow, then, the most basic resource that costs energy is the difference between a cobit and a qubit. This turns out to be a resource called the \emph{zerobit} (or co-cobit) \cite{harrow2009time, hayden2012weak, 2020CMaPh.374..369H}, which satisfies
\begin{align}\label{eq:qubitcobit}
 \text{1 qubit = 1 cobit + 1 zerobit}
\end{align}
and
\begin{align}\label{eq:cobitebit}
 \text{1 cobit = 1 ebit + 1 zerobit}.
\end{align}
The simplest definition of a zerobit is as the adjoint of the cobit isometry, namely the map
\begin{align*}
 \alpha \ket{0}_A\ket{0}_B + \beta \ket{1}_A\ket{1}_B \stackrel{\text{zerobit}}{\to} \alpha \ket{0}_B + \beta \ket{1}_B.
\end{align*}
Initially, Alice and Bob share a bipartite state that they know is in a superposition of $\ket{00}$ and $\ket{11}$. Sending a zerobit from Alice to Bob means coherently erasing Alice's part of the state, so that Bob ends up with the full state in his lab. Suppose, for example, that Alice wants to send a qubit using only a cobit and a zerobit. To do so, she first uses the cobit to send part of the state to Bob, and then uses the zerobit to erase the part she was left with
\begin{align*}
 \alpha \ket{0}_A + \beta \ket{1}_A \stackrel{\text{cobit}}{\to} \alpha \ket{0}_A\ket{0}_B + \beta \ket{1}_A\ket{1}_B \stackrel{\text{zerobit}}{\to} \alpha \ket{0}_B + \beta \ket{1}_B.
\end{align*}
It follows that the left-hand side of \eqref{eq:qubitcobit} is no greater than the right-hand side as a communication resource. 
% Similar protocols can be used to show the opposite inequality for \eqref{eq:qubitcobit} and both inequalities for \eqref{eq:cobitebit}. 

To send a zerobit and cobit using a qubit, Alice first implements a CNOT gate (with the cobit qubit $A_1$ acting as the control qubit and the zerobit qubit $A_2$ as the target) and then sends $A_2$ to Bob
\begin{align}
&\left(\alpha_1 \ket{0}_{A_1} + \beta_1 \ket{1}_{A_1}\right) \left(\alpha_2 \ket{00}_{A_2B_2} + \beta_2 \ket{11}_{A_2B_2}\right) \\&\qquad\nonumber\stackrel{CNOT_{A_1 A_2}}{\to} \frac{1}{\sqrt{2}}\left(\alpha_1 \ket{0}_{A_1}\left(\alpha_2 \ket{00}_{A_2 B_2} + \beta_2 \ket{11}_{A_2 B_2}\right) + \beta_1 \ket{1}_{A_1}\left(\alpha_2 \ket{10}_{A_2 B_2} +\beta_2 \ket{01}_{A_2 B_2}\right) \right) \\&\qquad\nonumber\stackrel{\text{qubit}_{A_2 \to B_1}}{\to} \frac{1}{\sqrt{2}}\left(\alpha_1 \ket{0}_{A_1}\left(\alpha_2 \ket{00}_{B_1 B_2} + \beta_2 \ket{11}_{B_1 B_2}\right) + \beta_1 \ket{1}_{A_1}\left(\alpha_2 \ket{10}_{B_1 B_2} + \beta_2 \ket{01}_{B_1 B_2}\right) \right) \\&\qquad\nonumber\stackrel{CNOT_{B_2 B_1}}{\to} \left(\alpha_1 \ket{00}_{A_1 B_1} + \beta_1 \ket{11}_{A_1 B_1}\right) \left(\alpha_2 \ket{0}_{B_2} + \beta_2 \ket{1}_{B_2}\right).
\end{align}
To send a cobit using an ebit and a zerobit, Alice applies a CZ gate to the cobit input $A_1$ and her half of the ebit $A_2$ and then sends $A_2$ to Bob as a zerobit 
\begin{align}
\frac{1}{\sqrt{2}}&\left(\alpha \ket{0}_{A_1} + \beta \ket{1}_{A_1}\right) \left(\ket{00}_{A_2 B} + \ket{11}_{A_2 B}\right) \\&\nonumber\stackrel{CZ_{A_1A_2}}{\to} \frac{1}{\sqrt{2}}\left(\alpha \ket{0}_{A_1}\left(\ket{00}_{A_2 B} + \ket{11}_{A_2 B}\right) + \beta \ket{1}_{A_1}\left(\ket{00}_{A_2 B} - \ket{11}_{A_2 B}\right) \right) \\&\nonumber\stackrel{\text{zerobit}_{A_2 \to B}}{\to} \alpha \ket{0}_{A_1}\ket{+}_{B} + \beta \ket{1}_{A_1}\ket{-}_B \stackrel{H_{B}}{\to} \alpha \ket{0}_{A_1}\ket{0}_{B} + \beta \ket{1}_{A_1}\ket{1}_B.
\end{align}
Finally to create an ebit and send a zerobit using a cobit, Alice applies a Hadamard gate to the zerobit input and then sends it using the cobit
\begin{align}\label{eq:protocol4}
\alpha \ket{0}_A\ket{0}_{B_1}& + \beta \ket{1}_A\ket{1}_{B_1} \stackrel{H_{A}}{\to} \alpha \ket{+}_A\ket{0}_{B_1} + \beta \ket{-}_A\ket{1}_{B_1}\\&\nonumber\stackrel{\text{cobit}_{A \to B_2}}{\to} \alpha \ket{0}_{B_1} \left(\ket{00}_{A B_2}+ \ket{11}_{AB_2}\right) + \beta \ket{1}_{B_1}\left(\ket{00}_{AB_2} - \ket{11}_{AB_2}\right)
\\&\nonumber\stackrel{CZ_{B_1 B_2}}{\to} \left(\alpha \ket{0}_{B_1} + \beta \ket{1}_{B_1}\right) \left(\ket{00}_{AB_2} + \ket{11}_{AB_2}\right).
\end{align}
% \GP{Too many equations here?} Collectively, the four protocols \eqref{eq:protocol1}-\eqref{eq:protocol4} lead to the resource equalities \eqref{eq:qubitcobit} and \eqref{eq:cobitebit}.

Comparing the two sides of \eqref{eq:qubitcobit}, we see that sending a zerobit into a black hole must have a minimum energy cost of $T_H \log 2$. In other words, the energy cost of throwing information into a black hole does not come from adding information to the interior; it comes from coherently erasing all record of that information from the black hole exterior. This is exactly what we should have expected from the generalized second law: it is the erasure of entropy from the exterior (not adding entropy to the interior) that needs to be compensated for by an increase in horizon area. From a purely exterior perspective, the idea that it is the erasure of information, or noise, that carries an energy cost is very familiar and is known as Landauer's principle. The novel feature of black holes is that this same principle is the limiting constraint on communication into the interior.

Indeed, while the resource theory we have just described seems fairly peculiar in the context of quantum communication, if we only focus on the exterior region it is also familiar. From an exterior perspective, cobit teleportation is just the depolarisation of a qubit into a maximally mixed state. 
In fact, the set of exterior state transitions that are possible without adding energy to a black hole are precisely those possible using so-called noisy operations. These are quantum channels $\mathcal{N}$ of the form
\begin{align*}
% \label{eq:noisyoperations}
 \mathcal{N}(\rho) = \Tr_E\left[U \left(\rho \otimes \frac{\mathds{1}_E}{d_E}\right) U^\dagger\right].
\end{align*}
Noisy operations define the resource theory of quantum purity \cite{horodecki2003local,2003PhRvA..67f2104H}: if the state $\rho_2$ can be written as
$ \rho_2 = \mathcal{N} (\rho_1)$
for some state $\rho_1$ and noisy operation $\mathcal{N}$, then we say that $\rho_1$ has more purity than $\rho_2$. In other words, purity is the resource that is valuable if you can only do noisy operations for ``free.''

In the usual contexts where purity is a relevant resource, the exterior is all we are interested in. The ``interior'' is whatever physical quantum system purifies the noisy operations in a Stinespring picture. Typically, however, this is some large, messy, complex environment that cannot be measured or controlled. (Indeed, if this was not the case, the environment would be a useful resource for purity and we would not be restricted to noisy operations!) As a result, it makes no sense to think about the resource theory of purity as a theory of communication from the system to the environment. From an exterior observer's perspective, a black hole is indeed a large, messy, complex environment that cannot be measured or controlled. However, for an interior observer, information that falls into a black hole is easy enough to access (albeit in a finite amount of time before hitting the singularity) while information that does not fall in is inaccessible. As a result, communication from the exterior to the interior is constrained by what one might call the ``Stinespring complement'' of the resource theory of purity, where communication channels are allowed if and only if the complementary channel is a noisy operation. 

What about communication protocols that do cost energy? In this case, the relevant exterior resource theory is that of quantum thermodynamics \cite{brandao2013resource, ng2019resource}. Noisy operations are replaced by thermal operations
\begin{align*}\label{eq:thermaloperations}
 \mathcal{N}(\rho) = \Tr_E\left[U \left(\rho \otimes e^{-\beta H_E} \right) U^\dagger\right].
\end{align*}
where $e^{-\beta H_E}$ is the state of a thermal bath (in this case the black hole) and the unitary $U$ must preserve total energy of the system and bath.\footnote{In the special case where $U$ does not move energy between the system and bath then we recover the resource theory of purity.} Entropy can now be dumped into the thermal bath, but only at the cost of removing energy from the system. The full resource theory of communication into the interior of a black hole (with energy use now allowed) is then the Stinespring complement of the resource theory of quantum thermodynamics. A communication channel is allowed only if its complementary channel is a thermal operation. This constraint underlies all the results described in this essay, from energy-free cobit teleportation to the energy cost of zerobits and energy extraction through ebits.

\section*{Conclusions}

In this essay, we explored the boundaries of the information/energy tradeoff in quantum black holes. We showed that, by using a novel decoherence protocol mediated by ``soft'' radiation, qubits can be teleported into a black hole without increasing the black hole's energy or entropy. The key mechanism was that, simultaneous with teleporting the qubit, the protocol generated a Bell pair of entanglement, which behaves as a synthetic Hawking quantum. Without this Bell pair, the protocol would violate both the generalized second law of thermodynamics and the unitarity of quantum gravity. This led us to leverage the powerful framework of quantum resource theories to demonstrate that the fundamental resource that carries an energy cost is not information entering the black hole, but ``zerobits'' -- the coherent erasure of information from the exterior. This is a sharp realization of Landauer's principle as a constraint on communication in quantum gravity.

\subsection*{Acknowledgments}
\noindent JKF is supported by the Marvin L.~Goldberger Member Fund at the Institute for Advanced Study and the National Science Foundation under Grant No. PHY-2207584. GP is supported by the Department of Energy through QuantISED Award DE-SC0019380 and an Early Career Award DE-FOA-0002563, by AFOSR award FA9550-22-1-0098 and by a Sloan Fellowship.

\bibliographystyle{JHEP.bst}
\bibliography{main}

\end{document}